\documentstyle[aps,prb,floats,epsfig]{revtex} 
\begin{document}
\draft
\twocolumn[\hsize\textwidth\columnwidth\hsize\csname
@twocolumnfalse\endcsname

\title{First-principles study of ferroelectric and antiferrodistortive
instabilities in tetragonal SrTiO$_3$  }
\author{Na Sai and David Vanderbilt}
\address{Department of Physics and Astronomy, 
Rutgers University, Piscataway, NJ 08854-8019}

\date{April 21, 2000}

\maketitle

\begin{abstract}

We carry out first-principles density-functional calculations of the
antiferrodistortive (AFD) and ferroelectric (FE) soft-mode instabilities
in tetragonal SrTiO$_3$, with the structural degrees of freedom
treated in a classical, zero-temperature framework.  In particular,
we use frozen-phonon calculations to make a careful study of the
anisotropy of the AFD and FE mode frequencies in the tetragonal
ground state, in which an $R$-point AFD soft phonon has condensed.
Because of the anharmonic couplings, the presence of this AFD
distortion substantially affects both the AFD and FE mode frequencies.
The AFD mode is found to be softer for rotations around a perpendicular axis
($E_{g}$ mode) than for rotations about the tetragonal axis ($A_{1g}$ mode), in agreement with experimental results.  The FE mode,
on the other hand, is found to be softer when polarized perpendicular
to the tetragonal axis ($E_u$ mode) than parallel to it ($A_{2u}$
mode).  The sign of this frequency splitting is consistent with
the experimentally reported anisotropy of the dielectric susceptibility
and other evidence.  Finally, we present a discussion of the influence
of various types of structural distortions on the FE instability and
its anisotropy.
\end{abstract}
\pacs{PACS numbers: 77.84.Dy, 77.80.Bh, 77.80.-e, 63.20.-e}
\vskip2pc]

\narrowtext

\marginparwidth 2.7in
\marginparsep 0.5in

\section{Introduction}

First-principles calculations are proving to be one of the most
powerful tools for carrying out theoretical studies of the
electronic and structural properties of materials.
A particularly successful application of this technique has been
its use in understanding the perovskite ferroelectric compounds.
These materials have important technological applications because
of their switchable macroscopic polarization and their piezoelectric
properties.  They are also attractive objects of fundamental
study because of the rich variety of phase diagrams that they display
as a function of temperature.   At high temperature, the ABO$_3$
perovskites retain full cubic symmetry. However various structural
phase transitions take place as the temperature is reduced.\cite{Lines}
For example, BaTiO$_3$ and KNbO$_3$ undergo phase transitions from
the cubic paraelectric (PE) phase to a succession 
of tetragonal, orthorhombic, and finally rhombohedral ferroelectric
(FE) phases.  In contrast, PbTiO$_3$ displays only a single transition,
from the cubic PE phase to a tetragonal FE phase.
In NaNbO$_3$ and PbZrO$_3$, non-polar antiferrodistortive (AFD)
or antiferroelectric (AFE) transitions take place,
associated with different types of tilts of the oxygen
octahedra, in addition to the FE transitions. It is understood
that the two types of transitions result from the condensation
of soft phonon modes at the Brillouin zone boundary with $q\neq0$ and
 at the zone center with $q=0$ (Ref.~\onlinecite{soft}).

The low-temperature behavior of SrTiO$_3$ has been an attractive
subject for experimental and theoretical study.  SrTiO$_3$ behaves
as an incipient ferroelectric\cite{Barrett} (similar to KTaO$_3$)
in the sense that it has a very large static dielectric response
and is only barely stabilized against the condensation of the FE
soft mode at low temperature.\cite{Lines,MW,SKK} As the temperature
is reduced, SrTiO$_3$ first undergoes a transition from the cubic
to a tetragonal AFD phase at 105K,\cite{FSW} but this transition is
of non-polar character and has little influence on the dielectric
properties.  The static dielectric response closely obeys a
Curie-Weiss law of the form of $\epsilon\sim (T-T_c)^{-1}$ at
temperatures above about 50K, but the divergence at a critical
temperature $T_c\sim$ 36K that would be expected from this formula
is not observed.\cite{MW,SKK,Viana}  Instead, the susceptibility
saturates at an enormous value of $\sim2\times10^4$ as $T$
approaches zero.  Because the system is so close to a ferroelectric
state, it is not surprising to find that it can be induced to become
ferroelectric, either by the application of electric field\cite{FSW},
uniaxial stress\cite{UweSakudo} or by the substitution of Ca ions on
the Sr sublattice.\cite{Bednorz}
Finally, the SrTiO$_3$ system also displays puzzling phonon
anomalies\cite{Courtens1,Courtens2,Scott,Sirenko} and
electrostrictive response\cite{Grupp} in the low-temperature
regime.

This peculiar behavior, especially the failure of the system to
condense into a FE phase at low $T$, has been the subject of
considerable theoretical study and speculation.
\cite{Chaves,Migoni,Muller2,Muller3,Martonak,Zhong96,Zhong95}
Recently, efforts have focused on the so-called ``quantum
paraelectric state'' postulated by M\"{u}ller and
Burkard,\cite{Muller2} who suggested that quantum fluctuations of
the atomic positions could suppress the FE transition and lead to a
stabilized paraelectric state.\cite{Muller2,Muller3,Martonak,Zhong96}
This hypothesis has received dramatic support from a recent
experiment showing that isotopically exchanged SrTi$^{18}$O$_3$
appears to become ferroelectric at 23K,\cite{Itoh} suggesting that
normal SrTi$^{16}$O$_3$ must be very close indeed to the
ferroelectric threshold.

First-principles calculations have already contributed significantly
to the understanding of the structural properties of SrTiO$_3$.
\cite{Zhong96,Zhong95,KingSmith,Lasota,Krakauer}  Calculations of
two groups\cite{Zhong95,KingSmith,Lasota} confirmed that SrTiO$_3$,
in its high-symmetry cubic structure at $T=0$, is unstable to both
FE and AFD distortions when the atomic coordinates are treated
classically.  Using classical Monte Carlo simulations on an
effective-Hamiltonian\cite{ZVR} fitted to the first-principles
calculations, Zhong and Vanderbilt\cite{Zhong95} predicted that
SrTiO$_3$ would first undergo the AFD transition at about
130 K, and then a further transition into a state with simultaneous
AFD and FE character at 70 K.  Anharmonic interactions between the
AFD and FE modes were found to be competitive, in the sense that
the presence of the AFD distortion was found to reduce the FE
transition temperature by about 20\%.  The same authors later showed
that, when a quantum-mechanical treatment of the atomic positions
was included via a quantum path-integral Monte Carlo simulation, the
AFD transition temperature was shifted very close to the experimental
one at 105 K, and the FE transition was suppressed down to the
lowest temperatures that could be studied ($\sim$5 K),\cite{Zhong96}
consistent with the experimental absence of a transition.
On the other hand, LaSota and coworkers have recently performed a
first-principles calculation of the ground state structural
properties and interactions between the FE and AFD instabilities in
SrTiO$_3$ using an LAPW approach.\cite{Krakauer,Private}
These authors found that the AFD tetragonal structure is stable against
the FE distortions, indicating no interaction between the AFD 
and FE modes. The conclusion is in contrast to the previous
theory.\cite{Zhong95} However, LaSota {\it et al.} made certain
approximations to the FE eigenmodes, as will be discussed below.

An additional motivation for a detailed theoretical study of SrTiO$_3$
is the opportunity to make contact with the remarkably systematic
experimental study of Uwe and Sakudo.\cite{UweSakudo}
These authors made careful measurements of the anisotropic dielectric
susceptibilities and Raman mode frequencies as a function of uniaxial
stress applied along different crystal orientations.
They also fitted their results, plus those of previous
experimental studies, to obtain a phenomenological description
of the couplings between the AFD, FE, and strain degrees of
freedom of the crystal.  In particular, their fit contains an
anisotropic coupling which, in the tetragonal AFD phase,
tends to favor FE distortions that are perpendicular to the
tetragonal axis over those that are parallel.
However, previous theoretical work has given an unclear picture of
this anisotropy.  On the one hand, Vanderbilt and Zhong \cite{Dhvferro}
found that the interaction between the FE and AFD modes, which was
mainly through the on-site anharmonic coupling, would tend to
favor FE modes polarized perpendicular to the AFD tetragonal axis,
in accord with experiment.  On the other hand, the Monte Carlo
calculations previously referred to\cite{Zhong95} indicated a
sequence of transitions with decreasing temperature in which
the FE order parameter of a $z$-polarized mode was found to develop
before the $x$ or $y$-polarized ones, indicating that
the $z$-polarized mode goes soft first.  Finally, the recent work
of LaSota {\it et al.} \cite{Krakauer,Private} suggests that there
is very little anisotropic coupling at all.
In view of these apparently conflicting theoretical results, we
felt it worthwhile to clarify this situation by carefully studying
the anisotropy of the FE distortion energy in the AFD ground
state.

With these motivations, we have carried out a thorough analysis
of the ground-state structural and dynamical properties of
tetragonal SrTiO$_3$ using first-principles density-functional
calculations.  This approach is based on a classical treatment
of the nuclear motions, and so is obviously unable to take into
account the quantum zero-point motion of the ionic positions
which becomes critically important at low temperature.
Nevertheless, from such a calculation one is still able to compute
interaction parameters for comparison with experiment, to identify
the effects which tend to suppress the FE instability in the
presence of the AFD state, and to obtain qualitatively a picture
of the dielectric anisotropy connected with the splitting of the
differently polarized FE modes in the AFD state.

The calculations are carried out using a plane-wave basis and ultrasoft
pseudopotentials.\cite{Vanderbilt90}  The theoretical equilibrium AFD
structure is obtained by minimizing the energy with respect to
cell volume, $c/a$ ratio, and internal parameters.  Frozen-phonon
calculations are then used to obtain the frequencies of
$\Gamma$-point and $R$-point phonon modes, including FE soft
modes, in the AFD ground state.  For this purpose, we make use
of a point-group symmetry analysis to reduce the complexity of
the distortions that need to be studied.  In order to interpret
the results in terms of a phenomenological description involving FE and
AFD mode distortions and strains, we use an approach similar to that
underlying the effective-Hamiltonian scheme first developed for the
BaTiO$_3$ system\cite{ZVR} and later applied to
SrTiO$_3$.\cite{Zhong96}  That is, we use the LDA calculations to
compute the values of the Taylor expansion coefficients of
the total energy with respect to these distortions,
and compare with the experimental determinations of
Uwe and Sakudo.\cite{UweSakudo}

The rest of the manuscript is organized as follows. In Sec.~II
we briefly describe the technique employed for the first-principles
calculations.  In Sec.~III we present and discuss the results of the
calculations.  We begin with the determination of the theoretical
tetragonal AFD structure, and then proceed to study the
energies of AFD and FE distortions about this reference structure,
with special attention to the anisotropies of the AFD and FE
mode frequencies.  Finally, we conclude in Sec. IV.

\section{Theoretical details}

Our {\it ab-initio} plane-wave pseudopotential calculations are
based on the Hohenberg-Kohn-Sham density-functional theory (DFT)
within the local-density approximation (LDA). Ultrasoft Vanderbilt
pseudopotentials \cite{Vanderbilt90} are used, with the O($2s$),
O($2p$), Ti($3s$), Ti($3p$), Ti($3d$), Ti($4s$), Sr($4s$),
Sr($4p$), and Sr($5s$) states included in the valence.  The
exchange-correlation energy is of the Ceperley-Alder form with
Perdew-Zunger parameterization.\cite{Ceperalder,Perdew}  A
conjugate-gradient minimization scheme\cite{KingSmith} is used
to minimize the Kohn-Sham energy, using a plane-wave cutoff of
30 Ry for all calculations.
Unless otherwise stated, our calculations are carried out at the
theoretical equilibrium lattice constant of 7.303\,a.u.\cite{KingSmith}
which is $\sim$1\% less than the experimental value of 7.365\,a.u., the
discrepancy representing the inherent LDA error.

\begin{figure}
\centerline{\epsfig{file=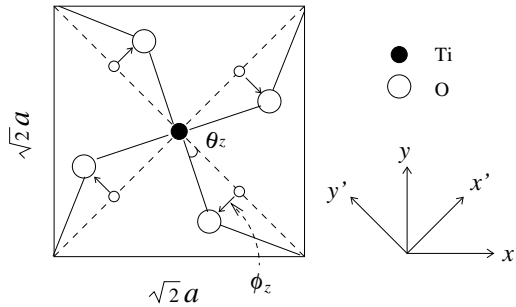,height=4cm}}
\vskip 0.5cm
\caption{Sketch of the TiO$_2$ layer in the unit cell of the
tetragonal AFD structure, illustrating rotation of octahedra
about the $z$ axis.  $x$-$y$ and $x'$-$y'$ are the coordinate
frames in the tetragonal cell and the original cubic cell,
respectively.}
\label{fig:rotation}
\end{figure}

Cubic SrTiO$_3$ has a simple cubic 5-atom unit cell with a common
lattice parameter $a$ along the [100], [010], and [001] directions.
We will briefly discuss some calculations carried out for a doubled
unit cell corresponding to the condensation of a soft AFD mode at
the (110)$\pi/a$ or $M$ point of the Brillouin zone (BZ) boundary.
However, most of our attention will be focused on the ground-state
tetragonal phase obtained by freezing in an AFD phonon mode at the
(111)$\pi/a$ or $R$ point of the BZ boundary.  This triply-degenerate
phonon mode corresponds to the rotation of the TiO$_6$ octahedra in
opposite directions from one cubic unit cell to the next,
followed by a small tetragonal strain. (Note that it is conventional to
label the phonon modes with respect to the simple-cubic BZ, even when 
they condense to lower the symmetry.)  Taking the rotation to be about 
the $z$ axis, we adopt a 10-atom tetragonal unit cell with lattice vectors
of length $\sqrt2a$, $\sqrt2a$, and $c$ along the [110],
[$\bar 1$10], and [001] directions, respectively.  (That is,
in our convention, $c/a$ is close to 1, not $1/\sqrt{2}$.) The
rotation of the oxygen atoms in the Ti$-$O plane is shown in
Fig.~\ref{fig:rotation}.
Throughout this paper, we will use $x'$ and $y'$ to denote the original
cubic directions ([100] and [010], respectively), while $x$ and $y$
are taken as parallel to the tetragonal lattice vectors along [110]
and [$\bar 1$10], respectively.  That is, the $x$-$y$ frame is rotated
by 45$^\circ$ relative to the $x'$-$y'$ frame ($z$ axes are congruent).

In all cases, we use a {\bf k}-point set that is equivalent to the
$6\times6\times6$ Monkhorst-pack mesh\cite{Monkhorst} in the BZ
of the simple cubic cell, corresponding to 108 {\bf k}-points in
the full BZ of the tetragonal cell.  The irreducible BZ then
contains 6 {\bf k}-points for the undistorted cubic structure;
10 {\bf k}-points for the tetragonal ground-state structure,
with or without additional $A_{1g}$ or $A_{2u}$ mode displacements;
and 20 {\bf k}-points for the tetragonal structure with additional
$E_u$ mode displacement.

\section{Results and Discussions}

\subsection{AFD instability in cubic unit cell}

\begin{figure}
\centerline{\epsfig{file=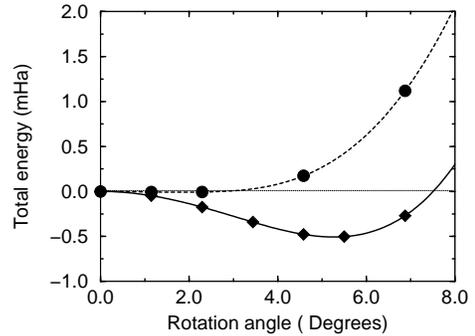,angle=270,width=6cm}}
\vskip 0.5cm
\caption{Calculated values (symbols) and fits (curves) of the
total energy per 10-atom cell as a function of octahedral rotation angle
$\theta_z$, computed at the frozen theoretical cubic lattice
constant $a$=7.303\,a.u., for both $R$-point (diamonds) and $M$-point
(circles) octahedral rotation modes.}
\label{fig:E-phi}
\end{figure}

To establish notation, we let the energy of an AFD phonon mode per 
10-atom cell of the cubic perovskite
structure be expanded up to fourth order in $\phi_z$,
\begin{equation}
E=E_0+\frac{1}{2}\kappa\phi_z^2+A_x\phi_z^4\;\;,
\label{E-phi}
\end{equation}
where $\phi_z=(a/2)\sin\theta_z$ is the magnitude of the oxygen-atom
displacement associated with the rotation of the oxygen octahedra
around the [001] axis, as shown in Fig.~\ref{fig:rotation}.
In Fig.~\ref{fig:E-phi} we show the computed values of the total
energy versus rotation angle for AFD modes at both the $M$ and
$R$-points of the BZ (corresponding to in-phase or out-of-phase 
rotations in neighboring planes of octahedra along $z$
respectively).  These were computed at zero strain, i.e., with the
lattice vectors fixed to be those of the theoretical equilibrium
cubic structure ($a$=7.303\,a.u.).  As can be seen, the computed total 
energy versus rotation angle can be fitted very well by the quartic 
Eq.~(\ref{E-phi}). Defining
the mode stiffness $\kappa$=$\partial^2E/\partial\phi_z^2$,
we find that $\kappa<0$ for both $M$- and $R$-point modes as
shown in Fig.~\ref{fig:E-phi}.  Nevertheless, the magnitude of
$\kappa$ for the $M$-point mode is only $\sim$10\% that of the
$R$-point mode, indicating that the instability at the $R$ point
is much stronger than that at the $M$ point.  Consequently, for
the remainder of this paper we will limit our discussion to
$R$-point distortions only.

\begin{figure}
\centerline{\epsfig{file=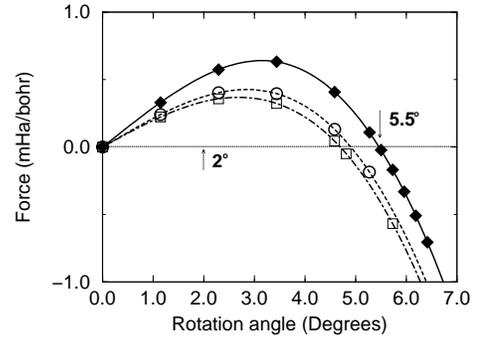,angle=270,width=6cm}}
\vskip 0.5cm
\caption{Calculated values (symbols) and fits (curves) of the
force experienced by an oxygen atom as a function of octahedral
rotation angle $\theta_z$, computed at the theoretical cubic
($a$=7.303\,a.u., diamonds), the experimental $T$=0 ($a$=7.365\,a.u.,
circles), and the experimental room-temperature ($a$=7.38\,a.u.,
squares) cubic lattice constants.  Arrows indicate the theoretical
and experimental equilibrium rotation angles of 5.5$^\circ$ and
2$^\circ$, respectively.}
\label{fig:F-phi}
\end{figure}

As can be seen from Figs.~\ref{fig:E-phi} and \ref{fig:F-phi},
the equilibrium octahedral rotation angle is found to be
$\theta_z$=5.5$^\circ$, significantly larger than the
zero-temperature experimental value of 2.1$^\circ$.\cite{Courten}
Since the theoretical equilibrium lattice constant (7.303\,a.u.) is
somewhat smaller than the experimental one, we also carried out
similar total-energy calculations at the extrapolated
zero-temperature (7.365\,a.u.) and room-temperature (7.38\,a.u.)
experimental lattice constants.\cite{Landolt}  The results shown
in Figure \ref{fig:F-phi} confirm that increasing the lattice
constant or crystal volume tends to suppress the AFD instability,
as expected from previous work.\cite{Zhong95,Zhong96}  However,
the resulting variation of the equilibrium rotational angle is
too small to explain the experimental observation, changing only
marginally to 4.89$^\circ$ and 4.69$^\circ$ at the zero- and
room-temperature experimental lattice constants, respectively.

These results demonstrate that the underestimate by $\sim1\%$ of the
lattice constant by the LDA is not the primary factor
responsible for the
theoretical overestimate of the rotation angle.  Moreover, we
shall see in the next subsection that the inclusion of strain
relaxation effects only acts to increase (slightly) the theoretical
equilibrium rotation angle.  Thus, we think that the smaller
observed value of the AFD rotation angle can most likely be
attributed to the quantum fluctuations associated with the motion
of the oxygen atoms.  This effect is not included in the theory,
and should act to reduce the amplitude of symmetry-breaking
distortions.  While previous work has indicated that the quantum
fluctuations should have a weaker effect on the AFD modes than
upon the FE ones,\cite{Zhong96} the effect on the AFD modes could
still be quite significant. An alternate possibility is simply
that the underestimate is a result of LDA error not associated
with the lattice constant.  In any case, we have chosen to complete
our theoretical investigations by considering distortions about
our theoretical ground-state AFD structure, keeping in mind that
the results should be interpreted with the overestimate of the
rotation angle in mind.

\subsection{AFD modes in the tetragonal structure}

To study the ground-state tetragonal structure, the lattice
strains also need to be taken into account.  We adopt the usual
Voigt notation $x_i$ for the strain tensor, but set $x_4=x_5=x_6=0$
because such off-diagonal shear strains will not enter into our
considerations.  We take $i$=1, 2, 3 corresponding to the $x'$,
$y'$, and $z$ pseudocubic axes for both strains $x_i$ and rotations
$\phi_i$.
Expanding the energy up to quartic order in AFD
amplitudes, quadratic order in strain, and leading order in the
strain-AFD coupling, the symmetry-allowed contributions are
\begin{eqnarray}
E=E_0&+&\frac{1}{2}\kappa\sum_{i}\phi_i^2+A_x\sum_{i}\phi_i^4
  +A^{n}_{x}\sum_{i<j}\phi_{i}^2\phi_{j}^2\nonumber\\
     &+& \frac{1}{2}c_{11}\sum_{i} x_{i}^2+c_{12}\sum_{i<j}x_{i}x_{j}\nonumber\\
     &-& b_{11}\sum_{i}x_i\phi_{i}^2-b_{12}\sum_{i<j}x_i\phi_{j}^2)\;\;.
\label{eq:F}
\end{eqnarray}
We choose the tetragonal ground state to be oriented along the
$z$-axis, with $x_1=x_2\ne0$, $\phi_1=\phi_2=0$, and $\phi_3\ne0$.
In this ground state, phonon modes corresponding to additional
oscillations of $\phi_i$ belong either to the $E_g$ ($i$=1 or 2)
or the $A_{1g}$ ($i$=3) representation of the tetragonal $D_{4h}$
point group.  For symmetry-preserving ($A_{1g}$) distortions, it is
convenient to re-express the three strain components in terms of a
volume strain $\overline{x}$ and a shear strain $v$ according to
\begin{eqnarray}  \begin{array}{ll}
x_1=x_2=\overline{x}-v \;\;,\\ 
\\
x_3=\overline{x}+2v\;\;. \end{array}
\end{eqnarray} 
In terms of these variables, the second and third lines of
Eq.~(\ref{eq:F}) can then be rewritten as
\begin{equation}
E^{\rm elastic}(\overline{x},v)=
   \frac{3}{2}\alpha\overline{x}^2+3\beta v^2
\end{equation}
and
\begin{equation}
E^{\rm coupling}({\overline{x},v},\phi_i)=
   -\eta\overline{x}\sum_{i}\phi^2_i-\gamma
   v(2\phi^2_3-\phi^2_1-\phi^2_2)\;\;,
\end{equation}
where
\begin{equation} \begin{array}{llll} \alpha=c_{11}+2c_{12}\;\;, \\ \\
\beta=c_{11}-c_{12} \;\;, \\ \\ \eta=b_{11}+2b_{12} \;\;, \\ \\
\gamma=b_{11}-b_{12} \;\;. \end{array} \end{equation}

To find the equilibrium rotation angle in the $A_{1g}$ ground state,
we hold $\phi_1$=$\phi_2$=0 and minimize $E$ in Eq.~(\ref{eq:F}) with
respect to $\overline{x}$ and $v$ at a fixed $\phi_3=\phi_z$. The
minimizing values are
\begin{equation} \begin{array}{ll}
\displaystyle \overline{x}^{\rm eq}=\frac{\eta}{3\alpha}\phi^2_z\;\;,\\
\\
\displaystyle v^{\rm eq}=\frac{\gamma}{3\beta}\phi^2_z \;\;. \end{array}
\end{equation}
Substituting into Eq.~(\ref{eq:F}),
\begin{equation}
E(\phi_z)=E_0+\frac{1}{2}\kappa\phi_z^2+A_X\phi_z^4 \;\;,
\end{equation}
where
\begin{equation}
\nonumber A_X=A_x-\frac{\eta^2}{6\alpha}-\frac{\gamma^2}{3\beta}\;\;.
\end{equation}
Thus, when the strain relaxation is taken into account, the equilibrium
rotation angle is given by
\begin{equation}
\phi_z^{\rm eq}=\sqrt{-\frac{\kappa}{4A_X}}\;\;.
\end{equation}

\begin{table}
\caption{Parameters in the effective Hamiltonian of SrTiO$_3$.  See
text.}
\begin{tabular}{cccc}
          &Uwe and Sakudo\tablenotemark[1]  & present   &Units\\ \hline
$\kappa$  &$-3.01\times10^{-6}$    &$-2.24\times10^{-5}$   & Ha/Bohr$^5$\\ 
$A_x$     &$ 5.16\times10^{-5}$   &$ 4.92\times10^{-5}$   & Ha/Bohr$^7$ \\ 
$A_X$     &$ 4.44\times10^{-5}$   &$ 4.14\times10^{-5}$   & Ha/Bohr$^7$ \\ 
$A^n_x$   &$ 2.13\times10^{-6}$   &$ 3.31\times10^{-6}$   & Ha/Bohr$^7$ \\ 
$A^n_X$   &$ 1.30\times10^{-5}$   &$ 2.30\times10^{-5}$   & Ha/Bohr$^7$ \\
$b_{11}$  &$ 1.23\times10^{-4}$   &$ 1.68\times10^{-4}$   & Ha/Bohr$^5$ \\ 
$b_{12}$  &$-2.37\times10^{-4}$   &$-2.70\times10^{-4}$   & Ha/Bohr$^5$  \\ 
$c_{11}$  &$ 1.14\times10^{-2}$   &$ 1.30\times10^{-2}$   & Ha/Bohr$^3$ \\ 
$c_{12}$  &$ 0.36\times10^{-2}$   &$ 0.33\times10^{-2}$   & Ha/Bohr$^3$ \\ 
$\phi_0$  &$2.0$                  &$5.5$                 & degree \\ 
\end{tabular}
\tablenotetext[1]{Ref.~\onlinecite{UweSakudo}}
\label{tab:coeff}
\end{table}

We determine all the interaction parameters $\kappa$, $A_x$, $A^n_x$,
$c_{11}$, $c_{12}$, $b_{11}$, and $b_{12}$ via a series of
finite-difference calculations of total energies and forces
within the LDA.  Table I lists our results and compares them
with the corresponding values determined by Uwe and Sakudo by fitting
to experiment \cite{UweSakudo} (all units have been converted to
atomic units).  We find very good agreement overall.\cite{explan-A}
The fact that $A_X<A_x$ implies that the inclusion of strain relaxations
strengthens the AFD instabilities at anharmonic order.
The equilibrium  rotation angle increases to 6.0$^\circ$ (compared with
5.5$^\circ$ for the cubic strain state), while the equilibrium
values of $\overline{x}$ and $v$ are found to be $-0.10\%$ and
$0.23\%$, respectively.

In the tetragonal AFD ground state, the frequencies of the soft
phonon modes associated with additional rotations of the oxygen
octahedra are given by evaluating
\begin{equation}
m_{\phi}\omega^2_i=\left.\frac{\partial^2E}{\partial\phi^2_i}\right|_{\rm eq}
\;\;,
\label{eq:omegadef}
\end{equation}
where $m_\phi=4m_{\rm O}$
is the mass factor associated with the oxygen rotational mode and
the derivative is to be evaluated under conditions of fixed strain
at the equilibrium structure (i.e., at the equilibrium values of
$\overline{x}$, $v$, and $\phi_z$). Eq.~\ref{eq:omegadef} gives
the frequencies of the $E_g$ and $A_{1g}$ modes to be, respectively, 
\begin{equation} \begin{array}{ll}
\omega^2_1=\omega^2_2=-\kappa(A^n_X/2m_{\phi}A_X)\;\;,\\
\\
\omega^2_3=-2\kappa A_x/m_{\phi}A_X  \;\;, \end{array}
\label{eq:omegaraw}
\end{equation}
where $A^n_X=A^n_x+\gamma^2/\beta$. The values of the two frequencies
are 45\,cm$^{-1}$ and 130.7\,cm$^{-1}$ respectively, so that the
frequencies of the $A_{1g}$ and $E_g$ modes are roughly in the ratio
3:1.  This is consistent with observed ratios of $\sim$2.5:1
in pressure-dependent experiments \cite{Ishidate} and $\sim$3:1 in
temperature-dependent experiments \cite{Slonczewski}. 

Up to this point, the analysis has been done at the theoretical equilibrium
lattice constant.  However, it is well known that the LDA tends to
underestimate the lattice constants of perovskites by $\sim$1\%.
\cite{KingSmith}  Moreover, past experience has shown that the
displacement patterns associated with soft modes may depend critically on
the lattice constant and strains.\cite{Cohen,Krakauer2}  To take these
effects into account, we adopted a strategy of applying a negative
hydrostatic pressure to the lattice to restore the experimental
lattice constant.\cite{ZVR}  Using Eq.~(\ref{eq:F}) and minimizing
the Gibbs free energy
\begin{equation}
G=E+3\overline{x}P\;\;
\end{equation}
with respect to $\overline{x}$ at fixed pressure $P$, we find
\begin{equation}
\overline{x}={\eta\over{3\alpha}}\phi_z^2-{P\over\alpha} \;\;,
\label{eq:xbar}
\end{equation}
while $v$ is independent of pressure. Then
\begin{equation}
G(\phi_z,P)={1\over2}\kappa_{\rm eff}\;\phi_z^2+A_X\phi_z^4-{3P^2\over2\alpha}
\end{equation}
where the effective harmonic coefficient is
\begin{eqnarray*}
\kappa_{\rm eff}=\kappa+2\eta{P\over\alpha}\;\;.
\end{eqnarray*}
Thus, at fixed pressure, the equilibrium rotation angle is
\begin{equation}
\phi_z^{\it eq}=\sqrt{-\frac{\kappa_{\rm eff}}{4A_X}}\;\;.
\label{eq:phiz}
\end{equation}

As one can see, the harmonic coefficient $\kappa_{\rm eff}$
depends upon the external pressure variable. It would thus be
possible, in principle, to adjust $P$ so as to fit the resulting rotation
angle to the experimental angle of 2.1$^\circ$. However,
the pressure needed to achieve this, $-$14.4GPa, would expand the
lattice constant to 7.48\,a.u., which is much larger than the experimental
value.  Instead, we adjust $P$ so as to fit the experimental lattice
constant.  That is, we adjust $P$ so that the volume strain is
$\overline{ x}=(a_{\rm exp}-a_{\rm theo})/a_{\rm theo}=0.849\%$, where
$a_{\rm exp}$=7.365\,a.u. and $a_{\rm theo}$ are the zero-temperature
experimental and theoretical lattice constants, respectively.
Substituting (\ref{eq:phiz}) into (\ref{eq:xbar}), we obtain
\begin{equation}
\overline{x}(P)=x_0-\frac{P}{\alpha_{\rm eff}}
\end{equation}
where
\begin{equation}  \begin{array}{ll}
\displaystyle x_0 = -\frac{\kappa\eta}{12\alpha A_X} \;\;,\\
\\
\displaystyle \alpha_{\rm eff}=
   \alpha\left[1+\frac{\eta^2}{6\alpha A_X}\right]^{-1}\;\;. \end{array}
\end{equation}
Inserting $\overline{x}=0.849\%$ leads to $P=-$5.26GPa.
The strains along the tetragonal and planar axes are found to be
1.135\% and 0.705\%, respectively. Relative to a cubic cell 
with the experimental lattice parameter 7.365\,a.u.,  the tetragonal cell we 
obtained is thus expanded along [001] while compressed along the [100]
and [010] directions. In this circumstance, the equilibrium rotation
angle is found to be 4.93$^\circ$.

Under these conditions, the $A_{1g}$ and $E_{g}$ soft-mode frequencies
of Eq.~(\ref{eq:omegaraw}) now become
\begin{equation} \label{eq:omega} \begin{array}{ll}
\omega^2_1=\omega^2_2=-\kappa_{\rm eff}(A^n_X/2m_{\phi}A_X)\;\;,\\
\\
\omega^2_3=-2\kappa_{\rm eff}A_x/m_{\phi}A_X \;\;. \end{array}
\end{equation}
From the coefficients in Table~\ref{tab:coeff},
$A_x/A_X$=$1.187$ and $2A^n_X/A_X$=$1.11$, so that clearly
$\vert\omega(E_g)\vert < \vert\omega(A_{1g})\vert$.
In fact, the two frequencies are
calculated to be $\omega(E_g)$=37\,cm$^{-1}$ and
$\omega(A_1g)$=109\,cm$^{-1}$.
(For comparison, $\omega(A_1g)$=124\,cm$^{-1}$ in
Ref.~\onlinecite{Private}, while the measured $\omega(E_g)$ and
$\omega(A_1g)$ in Ref.~\onlinecite{FSW} are 15\,cm$^{-1}$ and
48\,cm$^{-1}$ respectively).

To summarize the results so far, we have found that the AFD mode
condenses at the $R$ point of the cubic BZ, associated with a
triply-degenerate phonon of $\Gamma_{25}$ symmetry.  As a
consequence of the transition from the cubic to the tetragonal
state, the degenerate $R$-point modes split strongly into an $A_{1g}$
singlet and an $E_g$ doublet, the latter having a softer frequency
than the former.  This is in good qualitative agreement with
experiment.\cite{FSW}

The expansion approach used above for the AFD modes makes an
implicit assumption that the phonon eigenvectors from the cubic
structure are a good approximation to those in the tetragonal AFD
structure.  For the AFD modes, where the anisotropy is large, we do
not expect this approximation to be at all serious.  However, our
next task will be to analyze the FE mode anisotropy in the AFD
state.  As will be seen below, this turns out to be much more
delicate than for the AFD modes.  Thus, we have chosen to take a
different approach for the FE modes, in which the normal modes in
the AFD ground state are directly computed.  The symmetry analysis
needed to do this is given in the next subsection, and the FE
mode analysis is then given in the concluding subsections.

\subsection{Symmetry analysis of normal modes in AFD tetragonal phase}

In this section, we present some details of the point-group
symmetry analysis of the normal modes in the AFD tetragonal structure,
needed for the calculation of the frequencies of transverse optical
(FE) modes at the Brillouin zone center.

To harmonic order, the displacement energy can be expressed as
\begin{equation}
E=\frac{1}{2}\sum_{i,j,\alpha\beta}\Phi^{\alpha,\beta}_{i,j}\;
  u^\alpha_iu^\beta_j \;\;,
\end{equation}
where $u^\alpha_i$ is the displacement of sublattice $i$ in Cartesian
direction $\alpha$, and the force constant matrix $\Phi$ obeys the
symmetry conditions $\Phi^{\alpha,\beta}_{i,j}=\Phi^{\beta,\alpha}_{j,i}$
and $\sum_j \Phi_{ij}=0$.  The dynamical matrix is just related to the
force constant matrix by a diagonal mass tensor.

\begin{table}
\caption{Character table for point group D$_{4h}$.}
\begin{tabular}{lrrrrrrrrrr}
irreps&$E$&2$C_4$&$C_2$&2$C'_2$&$2C''_2$&$i$&$2S_4$&$\sigma_h$&2$\sigma_v$&2$\sigma_d$\\\hline
$A_{1g}$&1&1&1&1&1&1&1&1&1&1\\
$A_{2g}$&1&1&1&$-1$&1&1&1&1&$-1$&$-1$\\
$B_{1g}$&1&$-1$&1&1&1&$-1$&$-1$&1&1&$-1$\\
$B_{2g}$&1&$-1$&1&$-1$&1&$-1$&$-1$&1&$-1$&1\\
$E_{g}$&2&0&$-2$&0&2&0&0&$-2$&0&0\\
$A_{1u}$&1&1&1&1&$-1$&$-1$&$-1$&$-1$&$-1$&$-1$\\
$A_{2u}$&1&1&1&$-1$&$-1$&$-1$&$-1$&$-1$&1&1\\
$B_{1u}$&1&$-1$&1&$-1$&$-1$&1&1&$-1$&$-1$&1\\
$B_{2u}$&1&$-1$&1&1&$-1$&1&1&$-1$&1&$-1$\\
$E_{u}$&2&0&$-2$&0&0&$-2$&0&2&0&0\\
\end{tabular}
\end{table}

It is well known that the vibrational modes at a given $\bf k$-point in
the BZ of a crystal transform according to the corresponding
irreducible representations of the symmetry group for that
$\bf k$-point.  Such an analysis, which has previously been used to
construct the force-constant matrices for the FE modes in the cubic
perovskite structure \cite{KingSmith} and for all modes in the
tetragonal FE structure of PbTiO$_3$,\cite{Alberto} is applied
here to the zone-center modes in the AFD tetragonal structure of
SrTiO$_3$.  When SrTiO$_3$ condenses from the cubic into the tetragonal
AFD phase, the point group lowers from $O_h$ to $D_{4h}$.
Because we are interested in zone-center modes, the symmetry group
of $\bf k$ is just the $D_{4h}$ point group itself, which contains 16
symmetry operations that can all be generated from a fourfold rotation
$C_4$ and two mirror reflections $\sigma_h$ and $\sigma_d$. The character
table is shown in Table II. There are 10 irreducible representations
(irreps), of which two are two-dimensional. The AFD soft modes originating
from the cubic $\Gamma_{25}(R)$ phonons now belong either to the
$A_{1g}$ or $E_g$ irreps, depending on whether the octahedron rotation
axis is parallel or perpendicular to the tetragonal axis, respectively.
Similarly, the modes originating from cubic $\Gamma_{15}$ FE modes are
now either $A_{2u}$ or $E_u$, depending on whether the polarization is
parallel or perpendicular to the AFD axis.

\begin{table}
\caption{Symmetry analysis of the normal modes in tetragonal AFD structure.}
\begin{tabular}{lccccc}
irreps&dim.&Sr&Ti&O$_{1-2}$&O$_3$\\ \hline
$A_{1g}$&1 &          &           &$R$\tablenotemark[1]      & \\
$A_{2g}$&1 &          &           &$R$\tablenotemark[1]      &3$zR$ \\
$B_{1g}$&1 &$SzR$     &           &$R$\tablenotemark[1]      &    \\
$B_{2g}$&1 &          &           &$R$\tablenotemark[1]      &         \\
$E_{g}$ &2 &$S(x,y)R$ &           &$(E,O)zR$   &3$(x,y)R$  \\
$A_{1u}$&1 &          &$TzR$      &            &         \\
$A_{2u}$&1 &$Sz\Gamma$&$Tz\Gamma$ &$Ez\Gamma$  &$3z\Gamma$  \\
$B_{1u}$&1 &          &           &            &        \\
$B_{2u}$&1 &          &           &$Oz\Gamma$  &      \\
$E_{u}$ &2 &$S(x,y)\Gamma$&$T(x,y)(\Gamma)$&$E(x,y)\Gamma$&3$(x,y)\Gamma$\\
        &  &              &$T(x,y)(R)$     &$O(x,y)\Gamma$&              \\
\end{tabular}
\tablenotetext[1]{From decomposition of $(E,O)(x,y)R$.}
\end{table}

To label all of the displacement patterns associated with the
10-atom cell (30 degrees of freedom), we use the notation
$T\alpha K$ to denote a displacement associated with atom type $T$ in
pseudocubic Cartesian direction $\alpha$ and having ``phase relation''
$K$.  The five atoms types are abbreviated as `$S$' for Sr, `$T$' for Ti,
`3' for oxygen atoms making Ti-O chains in the $z$ direction, and
`1' and `2' for oxygen atoms in TiO$_2$ $x$-$y$ planes.  The
``phase relation'' $K$ is either `$\Gamma$' or `$R$' depending on whether
the two atoms of the same type in the 10-atom cell move in-phase
or out-of-phase (that is, whether they originate from $\Gamma$-point
or $R$-point modes of the parent cubic structure).
Note, however, that some individual displacements contribute to
more than one irrep (e.g., 1$z\Gamma$ and 2$z\Gamma$ contribute
to both the $A_{2u}$ and $B_{2u}$ irreps).  Thus, for the two
in-plane oxygens we introduce alternative ``type'' designations
$E$ (`even') and $O$ (`odd') in place of `1' and `2', where
\begin{equation} \begin{array}{ll}
\displaystyle u_{E \alpha K}= \frac{1}{\sqrt{2}} (u_{1 \alpha K}+u_{2 \alpha K})\;\;, \\
\\
\displaystyle u_{O \alpha K}= \frac{1}{\sqrt{2}} (u_{2 \alpha K}-u_{1 \alpha K})\;\;. \end{array}
\end{equation}
(six degrees of freedom for each atom type), classifying them according
to the irreducible representations to which they belong.
Note that the four ($E$,$O$)($x$,$y$)$R$ modes do not have a simple
one-to-one correspondence with the $A_{1g}$, $A_{2g}$, $B_{1g}$, and
$B_{2g}$ modes to which they give rise; these are indicated in the
table with just the notation $R$.

Since we have access to the Hellmann-Feynman forces \cite{Hellman} in
our first-principles ultrasoft-pseudopotential approach, it is
convenient to compute the force-constant matrix elements from
finite differences as
\begin{equation}
\Phi^{\alpha,\beta}_{i,j} = \frac{F^\alpha_i}{u^\beta_j}\;\;,
\end{equation}
where $F$ is the force that results from a sufficiently small
displacement $u$.  We can use the symmetry analysis to identify
the set of sublattice displacements that may participate in a given
normal mode, and to calculate the forces that arise at first order with
each such displacement.  For example, for the $A_{2u}$ FE mode,
we find the four displacements of types $Sz\Gamma$, $Tz\Gamma$,
$Ez\Gamma$, and $3z\Gamma$ may participate.  For each, a displacement
amplitude of 0.2\% of the lattice constant is chosen so that the harmonic
approximation is still well satisfied.  From each such calculation,
the resulting force vector is projected onto the same set of four
displacements, thus building up the $4\times4$ force-constant matrix.
This matrix is then symmetrized and diagonalized.  In a similar way,
the FE $E_u$ mode is represented in a 6$\times$6 subspace with basis
$Sx\Gamma$, $Tx\Gamma$, $TyR$, 3$x\Gamma$, $Ex\Gamma$, and $Oy\Gamma$.

\subsection{FE instability in the AFD tetragonal phase}

\begin{table}
\caption{Zone-center transverse optical phonon frequencies
in the cubic structure.  (All units in cm$^{-1}$.)}
\begin{tabular}{lrrr}
$\Gamma$point	&TO1	&TO2	&TO3\\ \hline
Current (7.303\,a.u.)& 42$\phantom{i}$&168	&549\\
Current (7.365\,a.u.)& 94$i$	&151	&521\\
PW\tablenotemark[1] (7.30\,a.u.)	&41$i$	&165	&546\\
LR\tablenotemark[2] (7.412\,a.u.)	&100$i$	&151	&522\\
Expt.\tablenotemark[3] 90K 	&42$\phantom{i}$ &175	&545 \\
\end{tabular}
\tablenotetext[1]{Ref.~\onlinecite{Zhong94}: Plane-wave pseudopotential method.}
\tablenotetext[2]{Ref.~\onlinecite{Lasota}: LAPW linear response method.}
\tablenotetext[3]{Ref.~\onlinecite{Servoin}: Fitted from experimental
infrared reflection spectra.}
\end{table}

Previous work of Zhong and Vanderbilt\cite{Zhong94} and LaSota
{\it et al.} \cite{Lasota} has indicated that cubic SrTiO$_3$, in
the absence of any AFD distortion, shows a FE instability.  We
have confirmed this result here by performing frozen-phonon
calculations and building up the $4\times4$ force-constant matrix
for $\Gamma_{15}$ modes polarized along $z$ (essentially the same
procedure outlined for the $A_{2u}$ modes at the end of the last
subsection, except performed for the cubic 5-atom cell).  The
non-zero eigenvalues of the corresponding dynamical matrix are
given in Table IV for both the theoretical equilibrium (7.303\,a.u.)
and expanded experimental (7.365\,a.u.) lattice constants, together
with other theoretical and experimental results for comparison.
Our results at the theoretical volume indicate that the FE soft
phonon mode frequency remains real, although it was imaginary
according to the earlier linear-response and plane-wave
calculations which used slightly different lattice
constants.\cite{Lasota,Zhong94}  However, at the zero-temperature
experimental volume, the frequency is found to be imaginary,
indicating that the FE instability is indeed very sensitive to
the crystal volume.  Increasing the crystal volume enhances
the FE instabilities, in agreement with experiment.\cite{Ishidate2,Zhong95}
This is in contrast to the AFD instability, which weakens with
increasing volume as shown in Sec.~III.A.  These results suggests
that there is an inherent competition between AFD and FE modes.

\begin{figure}
\centerline{\epsfig{file=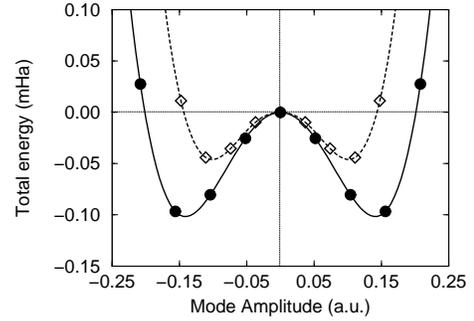,angle=270,width=6cm}}
\vskip 0.5cm
\caption{Total energy vs. the FE distortion amplitude at AFD tetragonal phase.
Circles and solid curve are for $E_u$ mode; diamonds and dashed 
curve are for $A_{2u}$ mode.}
\label{fig:Softmode}
\end{figure}

To understand the behavior of the FE modes in the zero-temperature
AFD tetragonal structure, we next perform frozen-phonon calculations
for this structure.  We use the AFD state described at the end of
Sec.~III.B, i.e., $P=-$5.26GPa, $V$ =798.989\,a.u.$^3$, $c/a =$ 1.004,
and $\phi_z$=4.93$^\circ$.  Both $E_u$ and $A_{2u}$ modes are
calculated using the method presented in Sec.~III.C.  Tables V and
VI present the resulting eigenfrequencies and eigenvectors for the
three lowest modes of each symmetry.  It can be seen that both the
$E_{u}$ and $A_{2u}$ soft-mode frequencies are imaginary (96$i$ and
90$i$\,cm$^{-1}$, respectively). Total-energy calculations are then
performed with the corresponding eigenmodes at different mode
amplitudes.  The double-well energy curves are as shown in
Fig.~\ref{fig:Softmode}.  One can see that the well depth of
the $E_{u}$ mode is substantially greater than that of the $A_{2u}$
mode, indicative of a much stronger instability for a distortion of
$E_{u}$ symmetry.

\begin{table}
\caption{Calculated eigenfrequencies and displacement patterns for the
A$_{2u}$ mode in tetragonal SrTiO$_3$.
Columns are labeled by mode eigenfrequency;
imaginary frequency indicates an unstable mode.}
\begin{tabular}{lddd}
		&~~~90$i$\,cm$^{-1}$ &~~~157\,cm$^{-1}$ &~~~515\,cm$^{-1}$ \\
\hline
$Sz\Gamma$ 		&$-$0.172	 &$-$0.700	&$-$0.036\\ 
$Tz\Gamma$ 		&$-$0.569	 &0.641		&0.058\\
$Ez\Gamma$ 		&0.636  	 &0.285 	&$-$0.583\\
3$z\Gamma$ 		&0.492   	 &0.128		&0.809\\
\end{tabular}
\end{table}

\begin{table}
\caption{Calculated eigenfrequencies and displacement patterns for the
E$_u$ mode in tetragonal SrTiO$_3$.  Columns are labeled by mode
eigenfrequency.}
\begin{tabular}{lddd}
 		&~~~96$i$\,cm$^{-1}$ &~~~240\,cm$^{-1}$ &~~~419\,cm$^{-1}$ \\ 
\hline
$Sx\Gamma$ 		&0.217  	&0.028		&0.013 \\
$Tx\Gamma$		&0.524		&$-$0.015	&$-$0.077 \\
$TyR$			&0.086		&0.014		&0.995 \\
$Ex\Gamma$		&$-$0.668	&$-$0.518	&0.055 \\
$Oy\Gamma$		&$-$0.034	&0.497		&$-$0.008 \\
3$x\Gamma$ 		&$-$0.472	&0.694		&0.024 \\
\end{tabular}
\end{table}

With respect to the FE mode frequencies, the comparison with experiment
is problematic because the theoretical values are imaginary (indicating
an instability) while the experimental values are not (consistent
with a non-FE $T$=0 ground state).  As indicated in the Introduction,
the experimental stabilization of the non-FE AFD structure is understood
to result from quantum fluctuations of the atomic coordinates.  It
should be emphasized that the purpose of the present calculations is to
study the low-temperature structural instabilities in a {\it classical}
framework, i.e., in the absence of quantum fluctuations.
Thus, the existence of the FE instabilities in our DFT-LDA ground
state calculations, taken together with earlier quantum Monte Carlo
simulations,\cite{Zhong96} tend to corroborate the hypothesis that the
quantum fluctuations are responsible for the experimental absence of a
real FE phase.

Nevertheless, we can make the following comparisons.
In the stress-induced Raman scattering measurement
of Uwe and Sakudo, the $E_u$ and $A_{2u}$ soft mode frequencies were
found to be 9.1$\pm$0.6\,cm$^{-1}$ and 19$\pm$1\,cm$^{-1}$ respectively
(this appreciable splitting between the two modes has frequently been
overlooked).\cite{UweSakudo}  The two squared frequencies produce a
difference of 3$\times$10$^2$\,cm$^{-2}$.  Our calculation shows the
difference between the two frequencies to be 6\,cm$^{-1}$, roughly the
same order of magnitude as the difference found experimentally.
However, the theoretical difference of squared frequencies is
10$\times$10$^2$\,cm$^{-2}$, or roughly three  times larger than the
experimental value. This is simply because of the large magnitude
of the imaginary frequencies.

In summary, we found imaginary frequencies for both the FE $E_u$
and $A_{2u}$ soft modes in the AFD tetragonal phase.
There is an apparent splitting between the two modes,
$\omega^2(E_u)< \omega^2(A_{2u})$,
suggesting that
the FE structure of $E_u$ symmetry is more energetically favorable
than the $A_{2u}$ one. This result is consistent with the fact that
the $A_{2u}$ FE mode is less easily observed in neutron
scattering experiments \cite{Shirane} because its energy is
higher than that of the $E_u$ phonon.\cite{UweSakudo}

\subsection{Influences of structural distortions on the stabilities}

In the previous subsection, we have calculated the FE phonon frequencies
for either a cubic structure in the absence of the AFD distortion,
or for a tetragonal AFD phase as observed experimentally.  To
understand the interaction of different distortions and their roles
in affecting the FE instabilities, we performed frozen-phonon
calculations for the cubic reference structure at the experimental
lattice constant both with and without the tetragonal strain, and
with and without the AFD rotation. In Table VII we present the
results for the FE phonon frequencies for each of these scenarios.

Observe that the two FE modes instabilities with different symmetries
depend on the strain distortion and AFD distortion in an opposite
sense.   For the $A_{2u}$ mode, a non-zero shear strain increases the FE
instability while an AFD rotation distortion reduces it.  The $E_u$
mode, however, depends on the two distortions in the opposite way.
Thus, the final sign of the frequency splitting (i.e., the sign of
the anisotropy) ultimately depends sensitively on a delicate partial
cancellation of the two contributions.  The final result is that
the $E_u$ mode is energetically slightly more unstable than the
$A_{2u}$ mode.

\begin{table}
\caption{Calculated $E_{u}$ and $A_{2u}$ phonon frequencies at different
distorted structures for SrTiO$_3$. $\overline{x}$, $v$ and $\theta_z$
denote uniform volume strain, shear strain, and rotation
angle about the $z$ axis, respectively.}
\begin{tabular}{cccrr}
\multicolumn{3}{c}{Distortions} &
\multicolumn{2}{c}{FE symmetry} \\
$\overline{x}$	&$v$	&$\theta_z$	&$A_{2u}$	&$E_{u}$\\ 
\hline
0	&0	&0		&42$\phantom{i}$&42$\phantom{i}$\\
0.00849	&0	&0		&94$i$		&93$i$	\\ 
0.00849	&0.00143&0		&112$i$		&83$i$	\\
0.00849	&0.00143&4.9$^\circ$	&90$i$		&96$i$	\\
0.00849	&0	&4.9$^\circ$	&63$i$		&107$i$	\\ 
\end{tabular}
\label{tab:aniso}
\end{table}

Finally, we have tested whether the use of the cubic eigen-mode
distortion in place of the true tetragonal one is a good
approximation for understanding the coupling of the FE and AFD
instabilities.  We calculated the expectation value
$\langle\xi_c\vert\Phi_T\vert\xi_c\rangle$ , where $\Phi_T$ is
the dynamical matrix in the tetragonal ground state structure (fourth
row of Table~\ref{tab:aniso}) and $\xi_c$ is the cubic eigen-mode
distortion vector.  The resulting imaginary frequencies are
85.4$i$ and 86.9$i$\,cm$^{-1}$ for the A$_{2u}$ and E$_u$ modes
respectively, compared with values of 90$i$ and 96$i$\,cm$^{-1}$
in Table~\ref{tab:aniso}. Thus, the splitting is much smaller when
the expectation value is used. This makes it clear that the nature of
the FE anisotropy is quite sensitive to the
tetragonal lattice strain associated with the AFD rotation distortion.
This suggests a probable explanation for the fact that the classical
MC simulations of Ref.~\onlinecite{Zhong96}, which are based on
use of a cubic mode eigenvector, predicted the anisotropy
incorrectly -- i.e., the $A_{2u}$ mode was found to go soft before
the $E_u$ one.

\section{Conclusion}

In this work, we have investigated both the AFD and FE structural
instabilities in the tetragonal phase of SrTiO$_3$. A unique aspect
of this work is that we have studied the FE frequencies carefully
using the exact eigenmode distortion obtained from the ground-state
tetragonal structure, whereas previous studies have made the
approximation of using the cubic eigenmode distortion
instead. We show that the instabilities have a sensitive dependence
on the crystal volume. We also found that the existence of the FE
instabilities are affected by the coupling to the shear strain and the
rotational distortion. Both types of distortion contribute, but with
opposite sign, so that it is a subtle cancellation between them
that determines the splitting of frequencies between the $E_u$ and
$A_{2u}$ modes.  The $E_u$ mode is found to be the more unstable
of the two in the ground-state AFD phase.

The degree of tilt of the oxygen octahedra is still overestimated in
our calculations, compared to the experimental results.  We attribute
this to the quantum fluctuations of the oxygen coordinates, which tend
to suppress the average AFD distortion amplitude in the experiments.
The AFD and FE instabilities show opposite trends with an increase of
crystal volume (weakening and strengthening, respectively).  However,
despite this competitive behavior, the FE instability is found to
coexist in the zero-temperature AFD phase.  Once again, quantum
zero-point fluctuations must be invoked to explain the experimental
observation that the AFD phase is stable against FE distortions.

\acknowledgments

Support for this work was provided by ONR Grant N00014-97-1-0048.




\begin{references}
\bibitem{explan-A} In the present work, off-diagonal shear strains are 
set to zero ($x_4=x_5=x_6=0$), and consequently the shear elastic
and coupling coefficients $b_{44}$ and $c_{44}$ are neglected, leading
to a discrepancy between our $A^n_X$ value and the value cited in
Ref.~\onlinecite{UweSakudo} in which the term $-b^2_{44}/c_{44}$
is added. However, our calculated values $A_x/A_X=1.187$ and
$2A^n_X/A_X=1.11$ are consistent with $A_x/A_X=1.1$ and
($2A^n_X+b^2_{44}/c_{44})/A_X=0.81$ respectively in
Ref.~\onlinecite{UweSakudo}.
\end{references}
\end{document}